# Title: Feasibility of a co-designed online nutrition education program for people with multiple sclerosis

Rebecca D. Russell[a*], Andrea Begley[a] Alison Daly[a] Eleanor Dunlop[a], Minh N. Pham[a], Lisa Grech[b], Lucinda J. Black[a,c]

[a]Curtin School of Population Health, Curtin University, Perth, Australia

[b]Department of Medicine at Monash Health, Faculty of Medicine, Nursing and Health Sciences, Monash University, Melbourne, Australia

[c]Institute for Physical Activity and Nutrition (IPAN), School of Exercise and Nutrition Sciences, Deakin University, Melbourne, Australia

*Corresponding author at
Dr Rebecca Russell
Curtin School of Population Health
Curtin University
GPO Box U1987,
Perth, WA 6845
Tel +61 8 9266 3160
Rebecca.Russell@curtin.edu.au

**Authorship**

RDR: Conceptualisation, Investigation, Writing – Original Draft, Writing – Review & Editing, Project administration, Funding Acquisition; AB: Conceptualisation, Methodology, Writing – Review & Editing, Supervision, Funding Acquisition; AD: Methodology, Formal analysis, Writing – Review & Editing, Funding Acquisition; ED: Resources, Writing – Review & Editing; MP: Data Curation, Formal analysis, Writing – Review & Editing; LG: Resources, Writing – Review & Editing; LJB: Conceptualisation, Methodology, Writing – Review & Editing, Supervision, Funding acquisition.

**Abstract**

**Objective:** Diet quality is important for people with multiple sclerosis (MS), but conflicting online information causes them confusion. People with MS want evidence-based MS-specific information to help them make healthy dietary changes, and we co-designed an asynchronous, online nutrition education program (Eating Well with MS) with the MS community. Our aim was to determine the feasibility of Eating Well with MS.

**Methods:** We used a single-arm pre-post design. The feasibility trial was a nine-week intervention with adults with confirmed MS. Feasibility outcomes: 1) demand (recruitment); 2) practicality (completion); 3) acceptability (Intrinsic Motivation Inventory: interest/enjoyment and value/usefulness subscales); and 4) limited efficacy testing (Diet Habits Questionnaire (DHQ); Critical Nutrition Literacy Tool (CNLT); Food Literacy Behaviour Checklist (FLBC)).

**Results:** The recruitment target (n=70) was reached. 87% completed at least one module and 57% completed the full program (five modules). The median interest/enjoyment rating was 5 out of 7 and median value/usefulness rating was 6 out of 7 (where 7 = 'very true'). Compared to pre-program, participants who completed any of the program had statistically significantly improved DHQ, CNLT, and FLBC scores.

**Conclusion:** Eating Well with MS was well received by the MS community and improved their dietary behaviours; demonstrating feasibility. Our findings support the use of co-design methods when developing resources to improve dietary behaviours.

**Keywords:** co-design, nutrition education, mixed-methods, online education, multiple sclerosis, diet

# 1.0 Introduction

Emerging evidence suggests that diet, a modifiable lifestyle factor, may influence the pathogenesis and course of multiple sclerosis (MS).[1] People with MS in Australia are advised by MS organisations to follow the Australian Dietary Guidelines to improve overall health, reduce the risk of vascular co-morbidities, and ensure adequate nutrient intake.[2, 3] These outcomes are important, because vascular co-morbidities are associated with increased disability progression[4, 5] and some nutrient deficiencies may accelerate demyelination.[6] High-quality diets, such as those in line with the Australian Dietary Guidelines, have been significantly associated with improved quality of life,[1, 7] lower relapse risk,[8] and reduced symptoms of depression[7, 9] in people with MS.

We have previously reported that 40% of Australians with MS make dietary modifications after diagnosis.[10] While some of the dietary changes aligned with a healthier diet (i.e., increased consumption of fruit and/or vegetables), some changes, such as eliminating all dairy foods, were not.[10] People with MS search the internet for information about diet,[11, 12] where they find conflicting information, including non-evidence-based restrictive diets that often promote eliminating entire food groups.[13] Our qualitative research has shown that conflicting dietary information causes angst when deciding what foods to choose/limit, and people with MS want MS-specific dietary advice.[14] Making dietary changes helps people with MS to feel more in control of their disease[12] and eHealth interventions can overcome physical and geographic barriers to participation and are well accepted by people with MS.[15]

Precipitated by our qualitative research with people with MS, we developed an online nutrition education program, Eating Well with MS (EWWMS), using co-design principles.[16, 17] Details of the co-designed program development have been published previously.[18] Briefly, EWWMS was underpinned by a theoretical framework (the self-determination theory[19] and the context, executive and operational systems (CEOS) theory[20]) and a program logic model. The program used 28 behaviour change techniques (BCTs) classified according to Michie et al.'s 93-item BCT Taxonomy,[21] including information about health consequences, behaviour practice/rehearsal, social comparison, and credible sources. To support the translation of nutrition information into dietary behaviour change, the program focussed on goal setting, action planning, and problem-solving. EWWMS was designed to provide participants with the knowledge and skills to: 1) manage their symptoms through healthy eating; 2) assess the quality of their eating habits; 3) select, prepare, and cook healthy meals; 4) judge the credibility of special diets that are marketed to people with MS; and 5) explain how researchers develop evidence in the field of nutrition and MS.

An important step in developing an evidence-based nutrition program is to run a feasibility study. A feasibility study is intended to generate evidence to determine if and how to scale up to an effectiveness study.[22] The aim of this study was to assess the feasibility of EWWMS. The domains used to assess feasibility were guided by the National Cancer Institute feasibility studies framework, namely demand (recruitment), practicality (completion, e.g., ability of participants to complete the intervention), acceptability, and limited efficacy testing.[22] The objectives were: 1) to determine the feasibility of EWWMS with respect to demand, practicality, and acceptability; and 2) to assess the efficacy of EWWMS with respect to diet quality, nutrition literacy, and food literacy.

## 2.0 Material and methods
### 2.1 Study design and participants

EWWMS was assessed using a single-arm pre-post design. A within subject study design was chosen to optimise statistical power, with each participant acting as their own control.[23] A power analysis showed that a sample of 35-40 participants would yield sufficient power to detect a modest change of 0.1 in the Diet Habits Questionnaire (DHQ) with a power of 0.8 with alpha at 0.05. The estimates of standard deviation and effect size used in the power calculation were based on a previous study using the DHQ as the outcome measure.[24] The aim was to recruit 70 participants to account for approximately 20% attrition between pre- and post-intervention measures.

This study was approved by [blinded for review] Human Research Ethics Committee (HREC2022-0020). Written informed consent was obtained from all participants.

This trial was registered with the Australian New Zealand Clinical Trials Registry https://www.anzctr.org.au/ Trial Id: ACTRN(12622000276752).

Participants were recruited between July-August 2022 through MSWA (a non-profit support and service provider to people living with neurological conditions in Western Australia) via newsletter and email, and our previous participant database. Potential participants were provided with the study details, screening process, time commitment, roles and responsibilities of the research team, and the rights of participants. The exclusion criteria were: aged less than 18 years; participation in the co-design workshops or interviews;[18] currently working with a dietitian in relation to a health condition; and having completed an online nutrition education program previously. Eligible participants could understand English language, had internet access, and provided online informed consent.

**2.2 Intervention**

Before the program commenced, participants were posted a package which contained hardcopy recipe booklets, the EWWMS Activity Book, and a brochure and poster on healthy eating.[3] The total program comprised seven modules (Table 1), with Modules 2-6 forming the core program. The program was asynchronous (self-paced, one module released each week for seven weeks) with two additional weeks allocated for program completion (intervention duration = nine weeks).

Table 1. Eating Well with MS weekly modules

| Module 1 | Welcome |
|---|---|
| Module 2* | Healthy eating is important for people with MS |
| Module 3* | Personalising your eating habits |
| Module 4* | Making changes to your eating habits |
| Module 5* | Understanding the diets that are marketed to people with MS |
| Module 6* | Understanding the research on diet and MS |
| Module 7 | Summary |

MS, multiple sclerosis. Average readability score of the text within the modules = Grade 8).
*Indicates modules that comprise the core program

**2.3 Data collection**

Participants were emailed weblinks to complete the baseline and post-intervention online surveys (Qualtrics, Provo, UT). Information collected at baseline included demographics (age, sex, highest level of education, employment status), disease characteristics (type of MS, date of diagnosis), and current dietary behaviours (if dietary changes were made for MS and/or other health conditions and if they were following a specific diet).

*2.3.1 Primary objective (feasibility)*

Demand was assessed using participant recruitment (were 70 participants recruited: yes/no), and length of time required for recruitment (were 70 participants recruited within six weeks: yes/no). Practicality was assessed using participant completion (the proportion of participants that completed each module and proportion that completed the core program). Acceptability was assessed using the interest/enjoyment and the value/usefulness subscales of the Intrinsic Motivation Inventory (IMI).[25] Both subscales contain seven questions assessed on a 7-point Likert scale ('not true at all' (1) to 'very true' (7)).

*2.3.2 Secondary objective (limited efficacy)*

Limited efficacy testing was measured using three tools. Dietary behaviour change was measured using the DHQ (20 questions pertaining to type of food and methods of food preparation), which has been previously modified for use by people withMS.[24] Each question scores from 1-5 and produces 8 dietary subscores: cereals; fruits and vegetables; omega-3 fatty acids; food choices; food preparation; takeaways and snacks; fat; and fibre. The overall score ranges from 20 to 100, where higher scores indicate better dietary habits.[24]

Nutrition and food literacy were considered determinants of dietary behaviour change. Nutrition literacy was measured using the Critical Nutrition Literacy Tool (CNLT; 19 questions: 8 questions in the engagement in dietary habits scale and 11 questions relating to taking a critical stance towards nutrition claims and their sources).[26] Each question is assessed on a 5-point Likert scale ('disagree strongly' (1) to 'agree strongly' (5)). Total scores ranged from 8-40 for engagement and 11-55 for nutrition claims, where higher scores indicate great critical nutrition literacy.[26] Food literacy was measured using the Food Literacy Behaviour Checklist (FLBC; 15 questions pertaining to frequency of behaviours in the last month in relation to planning, management, selection, preparing and eating food).[27] Each question is assessed using a 5-point Likert scale ('never' (1) to 'always' (5)). The total score ranges from 16-80, with higher scores indicating greater food literacy. [27]

Other measures that are known to impact on people's ability to change dietary behaviours were collected (disability, depression, anxiety, fatigue, and dietary stage of change). Disability was measured using the Patient-Determined Disease Steps (PDDS), a self-reported rating of disability ranging from 'normal' (0) to 'bedridden' (8).[28] Depression and anxiety symptoms were measuring using the Hospital Depression and Anxiety Scale (HADS), 7 questions about measuring depression symptoms and 7 questions measuring anxiety symptoms produce total scores ranging from 0 to 21 for each scale, where higher scores indicate greater symptoms of depression or anxiety.[29] Fatigue as measured by the Fatigue Severity Scale (FSS), two parts: a 9 questions which measure how fatigue affects activities, assessed using a 7-point Likert scale from 'strongly disagree' (1) to 'strongly agree' (7), and the Visual Analogue Fatigue Scale to assess global fatigue (1 question, score ranging from 'worst' (0) to 'normal' (10)), FSS total scores range from 9 to 63, with higher scores indicating greater fatigue severity.[30] Dietary stage of change was measured using the Healthy Dietary Stages of Change Instrument (HDSC), 4 questions relating to engagement in healthy dietary behaviours (e.g., "Do you currently engage in regular healthy eating") rated on a dichotomous "yes" or "no" and converted using a scoring algorithm to represent the degree of engagement along a 5-point continuum: 1 (precontemplation), 2 (contemplation), 3 (preparation), 4 (action), and 5 (maintenance).[31, 32]

### 2.4 Data Analysis

Pre- and post-intervention characteristics of participants were reported using frequency and percentage for categorical variables and mean and standard deviation (SD) or median and interquartile range (IQR) for continuous variables. Median (IQR) score was calculated for each subscale of the IMI. To assess possible covariates associated with the DHQ, we examined if PDDS, HADS, FFS, and HDSC were associated with the total DHQ or with any of the eight subscores. Backward stepwise general linear modelling was used with all covariates entered in the initial model: sex (male/female), age (years), education (up to year 12; Trade/Apprenticeship/TAFE/Diploma; Bachelor's degree; Postgraduate degree), employment status (employed; unemployed looking for work; retired; disability pension; homemaker; volunteering/other), type of MS (RRMS, PPMS, SPMS, PRMS, other), time since diagnosis (years), diet changed because of MS (yes/no), following a specific diet (yes/no), HDSC (precontemplation, contemplation, preparation, action, maintenance), PDDS (1-8), depression (0-15), anxiety (0-20), FSS (9-63), interest/enjoyment (20-49), value/usefulness (9-49), CNLT (2.5-4.4), and FLBC (1.9-4) scores. For limited efficacy testing, each covariate with $p>0.10$ was systematically removed and the final model was bootstrapped (500 replicates). Adjustments were made to the bootstrapped model until all remaining variables had a $p\leq0.1$. Statistically significant associations for the dietary predictor variables was defined as $p<0.05$. Stata version 18 (StataCorp, College Station, TX, USA) was used for all analyses.

## 3.0 Results

### 3.1 Participant characteristics

A total of 67 participants completed the baseline survey and 50 completed the post-intervention survey (Table 2). Most of the participants who completed the post-intervention survey were female (94%), had achieved a bachelor's degree or higher (54%), and were currently employed (60%). The median time since diagnosis was 9 years. A quarter (24%) were currently making dietary changes for their MS and 12% were currently following a specific diet.

**Table 2** Participant characteristics at baseline and post-intervention

|  | Pre-intervention (n=67) | Post-intervention (n=50) |
|---|---|---|
| Sex, n (%) |  |  |
|    Males | 3 (4.5) | 3 (6.0) |
|    Females | 64 (95.5) | 47 (94.0) |
| Age, median (IQR) | 48.0 (16.0) | 50.0 (14.0) |
| Highest education completed, n (%) |  |  |
|    Up to year 12 | 16 (23.9) | 10 (20.0) |
|    Trade/Apprenticeship/TAFE/Diploma | 17 (25.4) | 13 (26.0) |
|    Bachelor's degree | 21 (31.3) | 18 (36.0) |
|    Postgraduate degree | 13 (19.4) | 9 (18.0) |
| Employment status, n (%) |  |  |
| Employed | 42 (62.6) | 30 (60.0) |
| Unemployed looking for work | 4 (6.0) | 4 (8.0) |
| Retired | 7 (10.4) | 5 (10.0) |
| Disability pension | 6 (9.0) | 6 (12.0) |
| Homemaker | 5 (7.5) | 4 (8.0) |
| Volunteering/other | 3 (4.5) | 1 (2.0) |
| **Disease characteristics** | | |
| Type of MS, n (%) | | |
| Relapsing-remitting | 51 (76.1) | 37 (74.0) |
| Primary-progressive | 4 (6.0) | 2 (4.0) |
| Secondary-progressive | 5 (7.5) | 4 (8.0) |
| Progressive-relapsing | 6 (9.0) | 6 (12.0) |
| Other | 1 (1.5) | 1 (2.0) |
| Time since diagnosis, years, median (IQR) | 8.8 (13.3) | 9.3 (14.3) |
| **Dietary characteristics** | | |
| Currently making dietary changes for MS, n (%) | | |
| No | 50 (74.6) | 38 (76.0) |
| Yes | 17 (25.4) | 12 (24.0) |
| Currently following a specific diet, n (%) | | |
| No | 60 (89.6) | 44 (88.0) |
| Yes | 7 (10.4) | 6 (12.0) |

IQR, interquartile range; MS, multiple sclerosis; SD, standard deviation; TAFE, Technical and Further Education

### 3.2 Feasibility

Demand: One round of recruitment notices was distributed, with 108 participants completing the screening questionnaire within a six-week timeframe, of which 96 (89%) were eligible. Of those 96 eligible participants, 73 (76%) consented to enrol by providing a postal address and a letter from their neurologist or other health professional confirming MS diagnosis. A total of 67 participants completed the baseline survey (3 withdrew prior to program starting, due to personal reasons) (Figure 1).

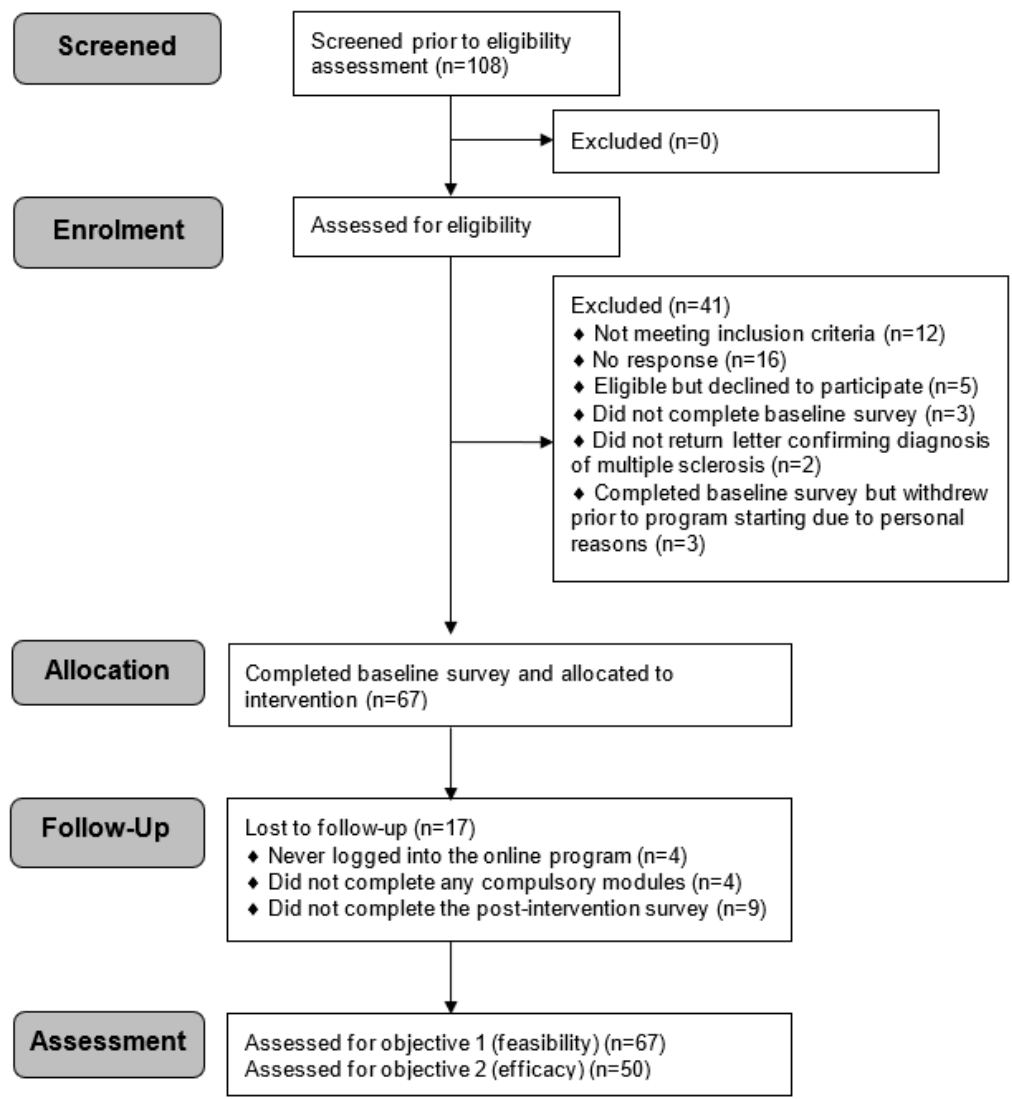

**Figure 1** CONSORT flow diagram[46] detailing participant screening and allocation.

A total of 59 (87%) participants completed at least one of the five core modules, 45 (67%) completed more than half, and 38 (57%) completed all five core modules of EWWMS within the nine-week timeframe. Figure 2 shows the percentages of participants who completed each of the core modules.

The median (IQR) interest/enjoyment rating was 5.0 (1.7) and median (IQR) value/usefulness rating was 5.9 (2.1) out of 7 (where 7 = 'very true').

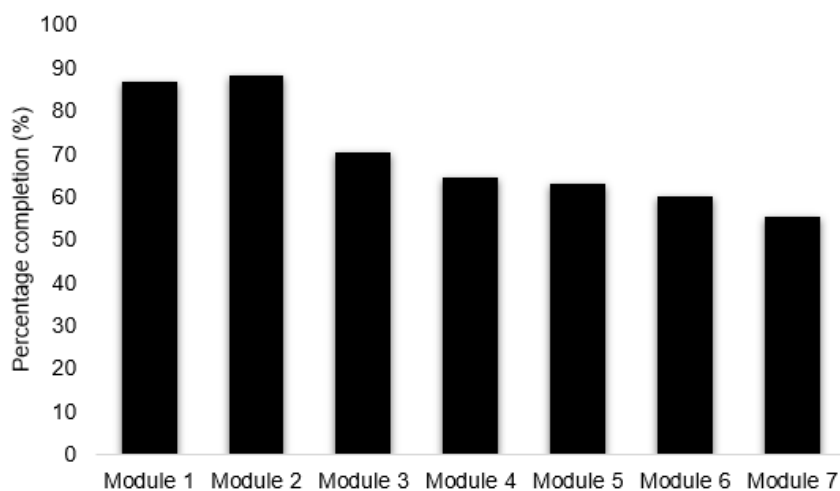

**Figure 2** Proportion of participants (n=67) that completed each of the modules in Eating Well with MS.

### 3.3 Limited efficacy testing

The total DHQ score and all the DHQ subscores (except the food preparation subscore) were statistically significantly higher at post-intervention compared to baseline (Table 3). Scores for both the CNLT and FLBC were statistically significantly higher at post-intervention compared to baseline. There were no other statistically significant differences in any of the psychosocial measures (disability, depression, anxiety, or fatigue symptoms, and dietary stage of change) between baseline and post-intervention.

**Table 3** Differences in scores between baseline and post-intervention

| Variable | Pre-intervention (n=67) | Post-intervention | $P$ |
|---|---|---|---|
| **DHQ total score,** mean (SD)* | 69.3 (10.1) | 75.9 (9.9) | <0.0001 |
| **DHQ subscores,** mean (SD)* | | | |
|    Cereals | 3.0 (0.8) | 3.6 (0.7) | <0.0001 |
|    Fruit and vegetables | 1.4 (0.5) | 1.7 (0.4) | <0.001 |
|    Omega-3 fatty acids | 2.6 (1.1) | 3.0 (1.3) | 0.002 |
|    Food choices | 3.2 (0.9) | 3.5 (0.9) | 0.002 |
|    Food preparation | 4.0 (0.7) | 4.0 (0.6) | 0.521 |
|    Takeaways and snacks | 3.0 (0.9) | 3.5 (1.0) | <0.0001 |
|    Fat | 3.4 (0.6) | 3.7 (0.6) | <0.001 |
|    Fibre | 3.0 (0.7) | 3.5 (0.6) | <0.0001 |
| **Nutrition literacy (CNTL),** mean (SD)* | 3.2 (0.4) | 3.5 (0.4) | <0.0001 |
| **Food literacy (FLBC),** mean (SD)† | 2.8 (0.5) | 3.0 (0.4) | <0.0001 |
| **Disability (PDDS),** median (IQR) [†] | 2.0 (3.0) | 2.0 (3.0) | 0.403 |
| **Depression (HADS),** median (IQR)[†] | 6.0 (5.0) | 7.0 (4.0) | 0.851 |
| **Anxiety (HADS),** median (IQR)[†] | 8.0 (6.0) | 8.0 (5.0) | 0.777 |
| **Fatigue (FSS),** median (IQR)[†] | 42.0 (30.0-54.0) | 44.0 (31.0-56.0) | 0.666 |
| **Dietary stage of change (HDSC),** n (%)[†] | | | 0.074 |
| Precontemplation | 0 (0.0) | 0 (0.0) | |
| Contemplation | 1 (2.0) | 3 (6.1) | |
| Preparation | 8 (16.3) | 2 (4.1) | |
| Action | 7 (14.3) | 11 (22.5) | |
| Maintenance | 33 (67.4) | 33 (67.4) | |

CNLT, Critical Nutrition Literacy Tool; DHQ, Diet Habits Questionnaire; FLBC, Food Literacy Behaviour Checklist; FSS, Fatigue Severity Scale; HADS, Hospital Depression and Anxiety Scale; HDSC, Healthy Dietary Stages of Change Instrument; PDDS, Patient-Determined Disease Steps; SD, standard deviation.
[*]n=50
[†]n=49

The only characteristics associated with change in DHQ score at post-intervention were fatigue and depression symptoms: when both the FSS and HADS-depression (HADS-D) scores were high there was a significant association with lower DHQ (Figure 3). Fatigue symptoms alone were not associated with lower DHQ.

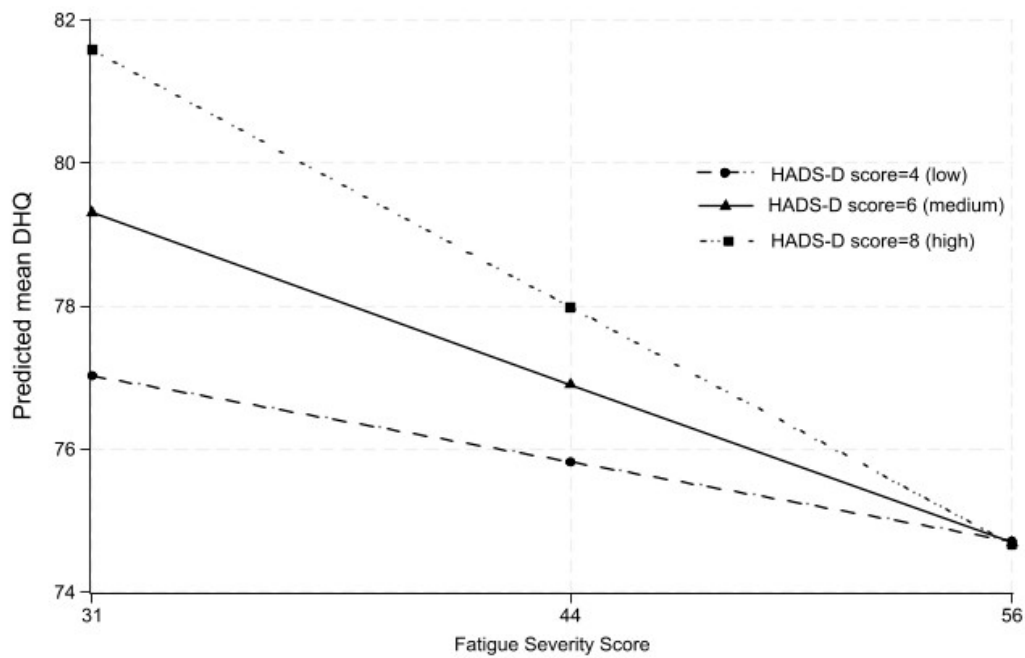

**Figure 3** Interactive effect of Fatigue Severity Score and Hospital Depression and Anxiety Scale Depression score (HADS-D) score on diet quality, measured using the Diet Habits Questionnaire (DHQ) post-intervention

## 4.0 Discussion and Conclusion

### 4.1 Discussion

There is growing consensus that improving the effectiveness of nutrition interventions relies on attention to design and feasibility/piloting.[33] Using the domains of demand (recruitment), practicality (completion), acceptability, and limited efficacy testing (dietary behaviours), we found that an asynchronous, online nutrition education program for people with MS consisting of seven modules, including five core education modules, was feasible.

There was sufficient demand to meet the recruitment target to detect a meaningful change in diet quality (measured using the DHQ) and it was feasible to request a letter from a neurologist/other health provider as evidence of MS diagnosis. The online nutrition education program was practical for people with MS to complete, with the completion rate (56%) exceeding the 40% completion rates of other online courses[34] and in alignment with other online MS programs.[35, 36] Our study completion rates highlight the practicality of the intervention. Additionally, the median interest/enjoyment and value/usefulness scores indicate that EWWMS was well-received, reinforcing the acceptability as an intervention for people with MS. The EWWMS program was codesigned with MS community input and best practice educational features, including positive tone and credible presenters, the success of which has also been shown in other acceptable online MS programs.[37-39]

The statistically significant improvements in the DHQ, CNLT, and FLBC scores demonstrate that EWWMS positively influenced the participants' dietary behaviours. A recent study that evaluated an online course to increase understanding about MS showed that 30% of participants with MS reported improved diet quality; however, no validated tool was used, only one module contained brief information about diet, and food literacy and nutrition literacy were not assessed.[35] Another study evaluating an online MS wellness program found no statistical difference in fruit and vegetable intake between the intervention and waitlist control groups.[40] Given the complexity of dietary behaviour change, it is not surprising that education programs that only aim to increase knowledge are considered ineffective for improving health-related outcomes in people with MS.[41] While we did not find any significant change in health-related outcomes between pre- and post-intervention (PDDS, HADS, FSS, and HDSC), we did identify participants who may be more vulnerable to lower diet quality - namely, people with higher fatigue and depression symptom scores. People with MS have identified the need for behavioural supports to encourage positive dietary change,[42] further justifying the importance of a theoretical framework and BCTs underpinning interventions to facilitate positive dietary behaviour change, and the need to evaluate other measures of dietary behaviours, such nutrition and food literacy.

EWWMS addressed the needs of people with MS for evidence-based and MS-specific dietary guidance, and the online format addressed potential physical and geographic barriers.[15] Epidemiological evidence suggests that a healthy diet can alleviate some MS symptoms, improve quality of life, and reduce relapse rate.[7-9] The positive changes in dietary behaviours demonstrated in this feasibility study suggest that interventions like EWWMS could contribute to improved well-being and disease management among people with MS. The majority (67%) of participants in this study indicated that they were already in the maintenance stage of dietary change (maintaining healthy dietary habits for the past six months and intending to continue[31]); however, some participants moved from preparation (not currently engaging in regular healthy eating, but intending to in the next 30 days) to action (currently engaging in regular healthy eating but for less than six months), highlighting the success of the program to prompt positive dietary change. While people with MS in Australia have reported making dietary changes towards a healthy diet,[10, 43] the mental effort required to make and sustain dietary changes[43] warrants the implementation of nutrition education programs such as EWWMS to support people with MS in making healthy dietary choices.

There are several strengths to this study. Firstly, the demographics of our participants aligned with those in other online MS programs (majority female[35, 40, 44, 45] and up to half with a university education[35, 45]). Secondly, we optimised statistical power by using participants as their own controls. Thirdly, we used recognised domains of feasibility, as guided by the National Cancer Institute feasibility studies framework.[22] Limitations of this study were: i) self-reported measures for assessing diet and psychosocial factors, introducing the potential for measurement error; ii) varying module completion rates and overall participant retention rate; and iii) no comparator arm; hence, we are not sure if the changes in dietary behaviours can be attributed to the intervention (although this was not the purpose of this feasibility study).

## 4.2 Conclusion

Our co-designed nutrition education program addressed the need for reliable, evidence-based nutrition education for people with MS and implementation was well-received, demonstrating feasibility with respect to the demand, practicality, acceptability, and limited efficacy testing. The potential for further tailoring of the content and delivery, investigating factors influencing engagement and strategies to enhance adherence to the program, and a larger-scale effectiveness study to validate the long-term impact of the program are warranted. Our findings support nutritionists and dietitians to use co-design methods when developing resources to improve dietary behaviours. Future research should also consider recruiting more participants in the preparation stage of change.

## 4.3 Practice Implications

Co-designing a nutrition intervention with the end-users resulted in a program that was well-received, suited to the needs of people with MS, and improved diet quality. Health professionals and researchers should collaborate and engage with the communities that they are developing health resources and interventions for, to increase the likelihood of acceptance and positive behaviour change.

We confirm all patient/personal identifiers have been removed or disguised so the patient/person(s) described are not identifiable and cannot be identified through the details of the story.

**Acknowledgements**
We thank the MS community for their involvement in co-designing EWWMS, and the participants who took part in this feasibility study.

**Funding**

This research was supported by MSWA and an MS Australia Incubator Grant (#21-1-072). LJB is supported by MSWA and an MS Australia Postdoctoral Fellowship. RR and AD are supported by MSWA. ED and LG are supported by MS Australia Postdoctoral Fellowships. MP was supported by the WA Department of Health Future Health Research and Innovation Fund. Funders had no role in the design, analysis, or writing of this article.

**Conflict of interest**

None

# 5.0 References

[1] Stoiloudis P, Kesidou E, Bakirtzis C, Sintila S-A, Konstantinidou N, Boziki M, et al. The role of diet and interventions on multiple sclerosis: a review. Nutrients 2022;14:1150. https://doi.org/10.3390/nu14061150.

[2] MS Research Australia. Modifiable Lifestyle Factors and MS. A Guide for Health Professionals, https://www.msaustralia.org.au/wp-content/uploads/2021/10/modifiable-lifestyle-factors-and-ms-a-guide-for-health-professionals.pdf; 2020 [accessed 10 June 2023].

[3] National Health and Medical Research Council [NHMRC]. Australian Dietary Guidelines, https://www.nhmrc.gov.au/guidelines-publications/n55; 2013 [accessed 10 June 2023].

[4] Marrie RA, Rudick R, Horwitz R, Cutter G, Tyry T, Campagnolo D, et al. Vascular comorbidity is associated with more rapid disability progression in multiple sclerosis. Neurology 2010;74:1041. https://doi.org/10.1212/WNL.0b013e3181d6b125.

[5] Kemp MC, Johannes C, van Rensburg SJ, Kidd M, Isaacs F, Kotze MJ, et al. Disability in multiple sclerosis is associated with vascular factors: An ultrasound study. J Med Imaging 2023;54:247-56. https://doi.org/10.1016/j.jmir.2022.11.017.

[6] Esposito S, Bonavita S, Sparaco M, Gallo A, Tedeschi G. The role of diet in multiple sclerosis: a review. Nutr Neurosci 2018;21:377-90. https://doi.org/10.1080/1028415X.2017.1303016.

[7] Marck CH, Probst Y, Chen J, Taylor B, van der Mei I. Dietary patterns and associations with health outcomes in Australian people with multiple sclerosis. Eur J Clin Nutr 2021;10.1038/s41430-021-00864-yhttps://doi.org/10.1038/s41430-021-00864-y.

[8] Simpson-Yap S, Oddy WH, Taylor B, Lucas RM, Black LJ, Ponsonby AL, et al. High Prudent diet factor score predicts lower relapse hazard in early multiple sclerosis. Mult Scler 2021;27:1112-24. https://doi.org/10.1177/1352458520943087.

[9] Saul A, Taylor BV, Blizzard L, Simpson-Yap S, Oddy WH, Probst YC, et al. Associations between diet quality and depression, anxiety, and fatigue in multiple sclerosis. Mult Scler Relat Disord 2022;63:103910. https://doi.org/10.1016/j.msard.2022.103910.

[10] Russell RD, Lucas RM, Brennan V, Sherriff JL, Begley A, The Ausimmune Investigator Group, et al. Reported changes in dietary behavior following a first clinical diagnosis of central nervous system demyelination. Front Neurol 2018;9:1-7. https://doi.org/10.3389/fneur.2018.00161.

[11] Riemann-Lorenz K, Eilers M, von Geldern G, Schulz K-H, Köpke S, Heesen C. Dietary interventions in multiple sclerosis: Development and pilot-testing of an evidence based patient education program. PLoS ONE 2016;11:e0165246. https://doi.org/10.1371/journal.pone.0165246.

[12] Russell RD, Black LJ, Sherriff JL, Begley A. Dietary responses to a multiple sclerosis diagnosis: a qualitative study. Eur J Clin Nutr 2018;73:601-8. https://doi.org/10.1038/s41430-018-0252-5.

[13] Beckett JM, Bird M-L, Pittaway JK, Ahuja KD. Diet and multiple sclerosis: scoping review of web-based recommendations. Interact J Med Res 2019;8:e10050-e. https://doi.org/10.2196/10050.

[14] Russell RD, Black LJ, Begley A. Navigating dietary advice for multiple sclerosis. Health Expect 2021;24:853-62. https://doi.org/10.1111/hex.13226.

[15] Scholz M, Haase R, Schriefer D, Voigt I, Ziemssen T. Electronic health interventions in the case of multiple sclerosis: from theory to practice. Brain Sci 2021;11:180. https://doi.org/10.3390/brainsci11020180.

[16] Slattery P, Saeri AK, Bragge P. Research co-design in health: a rapid overview of reviews. Health Res Policy Syst 2020;18:17. https://doi.org/10.1186/s12961-020-0528-9.

[17] Tay BSJ, Cox DN, Brinkworth GD, Davis A, Edney SM, Gwilt I, et al. Co-Design Practices in Diet and Nutrition Research: An Integrative Review. Nutrients 2021;13https://doi.org/10.3390/nu13103593.

[18] Russell R, Purdue J, Black LJ, Daly A, Begley A. A collaborative approach to designing an online nutrition education program for people with multiple sclerosis. Disabil Rehabil 2023;in press

[19] Deci EL, Ryan RM. Self-determination theory: a macrotheory of human motivation, development, and health. Can Psychol 2008;49:182-5. https://doi.org/10.1037/a0012801.

[20] Grech LB, Borland R. Using CEOS theory to inform the development of behaviour change implementation and maintenance initiatives for people with multiple sclerosis. Curr Psychol 2021;10.1007/s12144-021-02095-7https://doi.org/10.1007/s12144-021-02095-7.

[21] Michie S, Richardson M, Johnston M, Abraham C, Francis J, Hardeman W, et al. The behavior change technique taxonomy (v1) of 93 hierarchically clustered techniques: building an international consensus for the reporting of behavior change interventions. Ann Behav Med 2013;46:81-95. https://doi.org/10.1007/s12160-013-9486-6.

[22] Bowen DJ, Kreuter M, Spring B, Cofta-Woerpel L, Linnan L, Weiner D, et al. How we design feasibility studies. Am J Prev Med 2009;36:452-7. https://doi.org/10.1016/j.amepre.2009.02.002.

[23] Charness G, Gneezy U, Kuhn MA. Experimental methods: Between-subject and within-subject design. J Econ Behav Organ 2012;81:1-8. https://doi.org/10.1016/j.jebo.2011.08.009.

[24] Hadgkiss EJ, Jelinek GA, Weiland TJ, Pereira NG, Marck CH, van der Meer DM. The association of diet with quality of life, disability, and relapse rate in an international sample of people with multiple sclerosis. Nutr Neurosci 2015;18:125-36. https://doi.org/10.1179/1476830514Y.0000000117.

[25] Center for Self-Determination Theory. Intrinsic Motivation Inventory, https://selfdeterminationtheory.org/intrinsic-motivation-inventory/; [accessed 10 June 2023].

[26] Guttersrud O, Dalane JO, Pettersen S. Improving measurement in nutrition literacy research using Rasch modelling: examining construct validity of stage-specific 'critical nutrition literacy' scales. Public Health Nutr 2014;17:877-83. https://doi.org/10.1017/s1368980013000530.

[27] Paynter E, Begley A, Butcher LM, Dhaliwal SS. The validation and improvement of a food literacy behavior checklist for food literacy programs. Int J Environ Res Public Health 2021;18 https://doi.org/10.3390/ijerph182413282.

[28] Learmonth YC, Motl RW, Sandroff BM, Pula JH, Cadavid D. Validation of patient determined disease steps (PDDS) scale scores in persons with multiple sclerosis. BMC Neurology 2013;13:37. https://doi.org/10.1186/1471-2377-13-37.

[29] Wu Y, Levis B, Sun Y, He C, Krishnan A, Neupane D, et al. Accuracy of the Hospital Anxiety and Depression Scale Depression subscale (HADS-D) to screen for major depression: systematic review and individual participant data meta-analysis. BMJ 2021;373:n972. https://doi.org/10.1136/bmj.n972.

[30] Valko PO, Bassetti CL, Bloch KE, Held U, Baumann CR. Validation of the Fatigue Severity Scale in a Swiss cohort. Sleep 2008;31:1601-7. https://doi.org/10.1093/sleep/31.11.1601.

[31] Chiu C-Y, Lynch RT, Chan F, Rose L. The Health Action Process Approach as a motivational model of dietary self-management for people with multiple sclerosis: a path analysis. Rehabil Couns Bull 2012;56:48-61. https://doi.org/10.1177/0034355212440888.

[32] Raihan N, M. C. Stages of Change Theory. https://www.ncbi.nlm.nih.gov/books/NBK556005/Treasure Island (FL): StatPearls Publishing; 2023.

[33] Baker S, Auld G, Ammerman A, Lohse B, Serrano E, Wardlaw MK. Identification of a framework for best practices in nutrition education for low-income audiences. J Nutr Educ Behav 2020;52:546-52. https://doi.org/10.1016/j.jneb.2019.12.007.

[34] Reich J, Ruipérez-Valiente JA. The MOOC pivot. Science 2019;363:130-1. https://doi.org/10.1126/science.aav7958.

[35] Claflin SB, Mainsbridge C, Campbell J, Klekociuk S, Taylor BV. Self-reported behaviour change among multiple sclerosis community members and interested laypeople following participation in a free online course about multiple sclerosis. Health Promot J Austr 2021;10.1002/hpja.559https://doi.org/10.1002/hpja.559.

[36] Bevens W, Weiland TJ, Gray K, Neate SL, Nag N, Simpson-Yap S, et al. The feasibility of a web-based educational lifestyle program for people with multiple sclerosis: a randomized controlled trial. Front Public Health 2022;10:852214. https://doi.org/10.3389/fpubh.2022.852214.

[37] Bevens W, Reece J, Jelinek PL, Weiland TJ, Nag N, Simpson-Yap S, et al. The feasibility of an online educational lifestyle program for people with multiple sclerosis: A qualitative analysis of participant semi-structured interviews. Digit Health 2022;8:20552076221123713. https://doi.org/10.1177/20552076221123713.

[38] Claflin SB, Gates R, Maher M, Taylor BV. Building a successful Massive Open Online Course about multiple sclerosis: a process description. J Med Internet Res 2020;22:e16687. https://doi.org/10.2196/16687.

[39] Krause N, Riemann-Lorenz K, Rahn AC, Pöttgen J, Köpke S, Meyer B, et al. 'That would have been the perfect thing after diagnosis': development of a digital lifestyle management application in multiple sclerosis. Ther Adv Neurol Disord 2022;15:17562864221118729. https://doi.org/10.1177/17562864221118729.

[40] Titcomb TJ, Sherwood M, Ehlinger M, Saxby SM, Shemirani F, Eyck PT, et al. Evaluation of a web-based program for the adoption of wellness behaviors to self-manage fatigue and improve quality of life among people with multiple sclerosis: A randomized waitlist-control trial. Multiple Sclerosis and Related Disorders 2023;77:104858. https://doi.org/10.1016/j.msard.2023.104858.

[41] Claflin SB, Klekociuk S, Campbell JA, Bessing B, Palmer AJ, van der Mei I, et al. Association between MS-related knowledge, health literacy, self-efficacy, resilience, and quality of life in a large cohort of MS community members: A cross-sectional study. Mult Scler Relat Disord 2021;54 https://doi.org/10.1016/j.msard.2021.103158.

[42] Silveira SL, Richardson EV, Motl RW. Desired resources for changing diet among persons with multiple sclerosis: qualitative inquiry informing future dietary interventions. Int J MS Care 2021;24:175-83. https://doi.org/10.7224/1537-2073.2021-052.

[43] Weiss H, Russell RD, Black L, Begley A. Interpretations of healthy eating after a diagnosis of multiple sclerosis: a secondary qualitative analysis. Br Food J 2023;125:2918-30. https://doi.org/10.1108/BFJ-03-2022-0262.

[44] Dunne J, Chih HJ, Begley A, Daly A, Gerlach R, Schütze R, et al. A randomised controlled trial to test the feasibility of online mindfulness programs for people with multiple sclerosis. Mult Scler Relat Disord 2021;48:102728. https://doi.org/10.1016/j.msard.2020.102728.

[45] Cavalera C, Rovaris M, Mendozzi L, Pugnetti L, Garegnani M, Castelnuovo G, et al. Online meditation training for people with multiple sclerosis: A randomized controlled trial. Mult Scler 2019;25:610-7. https://doi.org/10.1177/1352458518761187.

[46] Moher D, Hopewell S, Schulz KF, Montori V, Gotzsche PC, Devereaux PJ, et al. CONSORT 2010 explanation and elaboration: updated guidelines for reporting parallel group randomised trials. BMJ 2010;340:c869. https://doi.org/10.1136/bmj.c869.